\documentclass[twoside]{article}
\usepackage{amsmath,amsfonts,bm,amssymb}
\usepackage[T2A]{fontenc}
\usepackage[cp866nav]{inputenc}
\usepackage{wrapfig}
\usepackage{cmpj}

\newcommand{\error}[1]{\small{#1}}
\newcommand{\best}[1]{\underline{#1}}

\newcommand{\Rn}{{{\cal R}_{1N}}}
\newcommand{\Rg}{{{\cal R}_{\mathrm{g}}}}
\newcommand{\Rh}{{{\cal R}_{\mathrm{h}}}}

\issue{2007}{10}{4(52)}{1} 

\title[Scaling of the DPD polymer chain]
{How does the scaling for the polymer chain in the dissipative
  particle dynamics hold?}
\author{J.M.Ilnytskyi\refaddr{a1,a3}, Yu.Holovatch\refaddr{a1,a2}}
\addresses{\addr{a1}Institute for Condensed Matter Physics of the National
Academy of
Sciences of Ukraine, \\1 Svientsitskii Str., 79011 Lviv, Ukraine%
\addr{a3}Institut f\"ur Physik, Universit\"at Potsdam, 
Am Neuen Palais 10, 14469 Potsdam, Deutschland
\addr{a2}Institut f\"ur Theoretische Physik, Johannes Kepler
Universit\"at Linz, 69
Altenbergerstr., 4040 Linz, Austria%
}
\date{Received October 9, 2007}

\begin{document}

\maketitle
\begin{abstract}
We performed a series of simulations for a linear polymer chain in a
solvent using dissipative particle dynamics to check the scaling
relations for the end-to-end distance, radius of gyration and
hydrodynamic radius in three dimensions. The polymer chains of up to
$80$ beads in explicit solvent of various quality are studied.  To
extract the scaling exponent $\nu$, the data are analyzed using linear
fits, correction-to-scaling forms and analytical fits to the
histograms of radius of gyration distribution. For certain
combinations of the polymer characteristics and solvent quality, the
correction-to-scaling terms are found to be essential while for the
others these are negligibly small. In each particular case the final
value for the exponent $\nu$ was chosen according to the best
least-squares fit. The values of $\nu$ obtained in this way are found
within the interval $\nu=0.55\div0.61$ but are concentrated mostly
around $0.59$, which is very close to the best known theoretical
result $\nu=0.588$. The existence of this interval is attributed both
to the peculiarities of the method and to the moderate chain lengths
being simulated. Within this shortcoming, the polymer chain in this
kind of modeling is found to satisfy the scaling relations for all
three radii being considered


\pacs  61.25.Hq, 61.20.Ja, 89.75.Da

\end{abstract}

\section{Introduction}\label{sec1}

Variety of physical, chemical, and biological phenomena which involve
polymers as well as variety of polymer characteristics one is
interested in have been naturally reflected in numerous theoretical
methods and models used for polymer description
\cite{desCloizeaux90}. In particular, it is generally recognized by
now, that the scaling properties of a long flexible polymer chain
immersed in a good solvent are perfectly described by the model of
self-avoiding walks (SAW) \cite{deGennes79}. A textbook example is
given by the mean square end-to-end distance $R_{1N}$ of a SAW of $N$
steps and that of a linear polymer chain, either of which scales in an
asymptotic limit $N \to \infty$ as:
\begin{equation} \label{1}
\langle R_{1N}^2 \rangle \sim N^{2\nu},
\end{equation}
with a universal exponent $\nu$. The value of the exponent depends on
the space dimension only and is the same for all SAW on different
three dimensional (3d) lattices. Relation (\ref{1}) is violated for
the finite $N$ and an approach to the asymptotics is governed by the
corrections to scaling. With an account of the first correction
equation~(\ref{1}) reads:
\begin{equation} \label{2}
\langle R_{1N}^2 \rangle = A N^{2\nu}\left(1 + \frac{B}{N^{\Delta}}+ \cdots\right).
\end{equation}
Here $A$, $B$ are non-universal amplitudes and $\Delta$ is an
universal correction-to-scaling exponent. Theoretical estimates for
the exponents follow from the field theoretical renormalization group
calculations \cite{Guida98}:
\begin{equation} \label{3}
\nu(3d)=0.5882 \pm 0.0011, \hspace{2em} \Delta(3d)=0.478 \pm 0.010.
\end{equation}

Whereas the SAW on the discrete lattice is a perfect model to study
scaling of long flexible polymer chains in a good solvent
\cite{Sokal95}, for obvious reasons it fails to describe most of the
other properties of the chain \cite{desCloizeaux90,Schaefer99}. A more
realistic model would be defined in a continuous space and involves
both monomer-monomer and monomer-solvent interactions. Brownian
dynamics is one of the candidates, as it operates on the mesoscale and
incorporates the random force acting on each monomer which mimics the
effect of a solvent \cite{FrenkSmit}. However, the method lacks
correct hydrodynamics limit. Similar but more rigorous technique,
i.e., the dissipative particles dynamics (DPD), was introduced by
Hoogerbrugge and Koelman \cite{HoogKoel} and later refined by
Espa\~{n}ol and Warren to satisfy the detailed balance
\cite{EspWarr}. The method has correct hydrodynamics limit
\cite{DPDHydro} and quickly gained much attention as a powerful
technique for mesoscale simulations of various kinds of macromolecules
\cite{FrenkSmit}.

Therefore, an important issue of analysis is to check how the scaling
laws (e.g. of equation~(\ref{1}) form) hold in the case of the
DPD-based simulations. This issue was partly considered in
\cite{Schlijp,Kong,Spenl,Symeon,Jiang}. However, some issues remain
unclear (see the next section \ref{sec2} for details). We see a good
reason to discuss in detail the scaling laws for the polymer chain in
a solvent in the DPD method. As far as the method was successfully
used for the simulations of branched, amphiphilic and other complex
molecules, the validation of correct scaling laws on such a
well-studied system as linear chain polymer would be a strong
supporting factor for extending the subject to the scaling laws in
more complex systems (e.g.  star-like, branched, dendritic systems,
etc.).

The set-up of our paper is as follows: in the next section \ref{sec2}
we give a brief review of former simulation results, section
\ref{sec3} contains the details of our simulations. Our main results
are presented in section \ref{sec4}, section \ref{sec5} gives
conclusions and outlook.

\section{Previous simulations}\label{sec2}

A number of various simulation techniques have been used to
numerically verify the scaling laws of the polymer in solution, namely
the molecular dynamics (MD), Monte Carlo (MC), DPD and others.  Here
we shall provide a rather brief account of references relevant to the
present study.

By using the MC method enriched by special techniques, one can move
very efficiently through the conformational space (see,
e.g. \cite{FrenkSmit}). As the result, the method is quite suitable
for the simulations of random walks (RW) and SAW
\cite{Grassberger93,LiMadr}. Li et al. \cite{LiMadr} used a pivot
algorithm and other techniques applied to the chain of $N=100\div1000$
monomers. The value for the scaling exponent obtained is $\nu=0.588$
and it agrees very well with the theoretical estimate (\ref{3}). To
study dynamical properties, however, one should turn to true dynamical
methods. Pierleone and Ryckaert \cite{PierRyk} performed a
comprehensive analysis of the scaling laws for both static (radius of
gyration, $R_{\mathrm{g}}$ and end-to-end distance $R_{1N}$) and
dynamical (diffusion coefficient, $D$) properties using MD
simulations. One of the important outcomes is that the static
properties are unaffected by the finiteness of the box size $L$
providing that $L/R_{\mathrm{g}}>3$. The scaling exponent values
$\nu=0.59$ and $0.568$ (for the $R_{1N}$ and $R_{\mathrm{g}}$,
respectively) were obtained. On the contrary, dynamical properties
were found to be strongly dependent on $L$. This fact was explained by
hydrodynamic interactions of the chain with its own images due to the
periodic boundary conditions (PBC). If, following Dunweg and
Kremer\cite{DunKrem} one uses the Kirkwood formula, then a correct
dynamical scaling is also recovered \cite{PierRyk}.

A number of simulations of linear polymers using DPD technique are
available. Schlijper et al.  \cite{Schlijp} were the first to study
static and dynamic scaling laws for the DPD chain in athermal
solvent. Later their results for this case were confirmed and extended
to include the solvents of variable quality by Kong et
al. \cite{Kong}. These authors studied $R_{\mathrm{g}}$ (among the
other properties) and have found for the athermal solvent $\nu=0.52$
which is closer to the polymer at the $\theta$-condition. With the
increase of solvent quality, the value $\nu\approx0.6$ close to a SAW
exponent was recovered. This problem was further addressed by Spenley
\cite{Spenl} who studied the $R_{1N}$ in polymer melt and solution.
In the latter case, relevant to our study, the exponent $\nu=0.58$ was
obtained. The previous result $\nu=0.52$ by Kong et al.\cite{Kong} was
commented as being influenced by the type of spring potential used
there. In our opinion, this discrepancy could also be due to
insufficient statistics.

It has also been debated whether the soft repulsion inherent to the
DPD method is sufficient to model the self-avoidance of the chains. In
particular, Symeonidis et al. \cite{Symeon} used additional
intramolecular interactions (e.g. Lennard-Jones-like repulsion term)
to increase the chain self-avoidance and experimented with different
forms of the bond springs (i.e. FENE, Hookean and WLC). One of the
conclusions made by the authors is, that the presence of FENE
potential alone ensures a correct value for the exponent
$\nu\approx0.6$ and no extra interactions for the improvement of
self-avoidance are required.

\looseness=-1Jiang et al. \cite{Jiang} have carried out an extensive
study of the hydrodynamical properties obtained via the DPD method. In
particular, scaling laws for both $R_{\mathrm{g}}$ and $R_{1N}$ for
the dilute athermal solution are relevant to our study. The authors
found the values for the scaling exponent $\nu$ weakly dependent on
the box size with the averages of $\nu=0.58$ for the radius of
gyration and $\nu=0.595$ for the end-to-end distance\cite{Jiang} (the
box sizes of $L=15\div30$ DPD length units were used and a range of
chain lengths was $N=10\div100$ DPD beads). The other conclusion given
in \cite{Jiang} is the one already mentioned above, claiming that the
original soft interactions are sufficient for self-avoidance of the
chains in the DPD method. Similarly to the MD study of Pierleone and
Ryckaert \cite{PierRyk}, the diffusion coefficient was found to be
highly sensitive to the simulational box size. However, after the
appropriate corrections to the system size were introduced, the
scaling law for the diffusion coefficient $D\sim N^{-\nu}$ was
obtained. Since, according to Kirkwood theory, the diffusion
coefficient $D$ is related to the hydrodynamic radius $R_{\mathrm{h}}$
\cite{DunKrem}, the latter is also shown to obey the correct scaling
law.

Despite the availability of simulational studies that concentrate on
static and dynamical scaling of the polymer in DPD method, there are
still certain points that need to be clarified, namely:
\begin{itemize}
\item[(i)] do all the relevant radii, $R_{1N}$, $R_{\mathrm{g}}$ and
      $R_{\mathrm{h}}$ scale with the same scaling exponent $\nu$?

\item[(ii)] are the corrections to scaling relevant and can these
      improve the results or make these more consistent?

\item[(iii)] in case of a good solvent, will the changes in the
      polymer-solvent interaction potential lead to any detectable
      drift of the scaling exponent $\nu$?
\end{itemize}

\noindent These three points will be addressed in the current study.

\section{Simulational details}\label{sec3}

\begin{figure}[th]
\centerline{\includegraphics[height=40mm,angle=00]{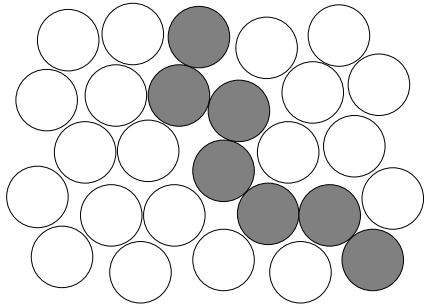}}
\caption{The polymer molecule in a solvent, as modeled in the DPD
method. Both monomers constituting a polymer chain and solvent
molecules are considered as interacting soft beads (grey and hollow
discs, correspondingly).
   \label{dpd}}
\end{figure}

In our study we closely follow the DPD method as discussed by Groot
and Warren \cite{GrWarr}.  Polymer molecule (see, figure~\ref{dpd}) is
modeled as a chain of soft beads connected via harmonic springs. The
bonding force acting on the $i$-th bead from its bond neighbour $j$
is:
\begin{equation}
  \vec{F}^B_{ij} = -kr_{ij}\hat{r}_{ij}\,,
\end{equation}
where $r_{ij}=|\vec{r}_{ij}|$, $\vec{r}_{ij}=\vec{r}_i-\vec{r}_j$,
$\hat{r}_{ij}=\vec{r}_{ij}/r_{ij}$ and $k=4$ is the spring
constant. Here and thereafter, the length, mass, time and energy
($k_{\mathrm{B}}T$) units are chosen to be equal to unity.
The solvent is described in an explicit way, in a form of isolated
beads of the same origin as the polymer ones (see,
figure~\ref{dpd}). Each $i$-th bead (polymer or solvent one) is
subject to a pairwise non-bonded force from its $j$-th
counterpart. The force acting on the $i$-th bead due to the
interaction with the $j$-th bead consists of three contributions:
\begin{equation}
  \vec{F}_{ij} = \vec{F}^{\mathrm{C}}_{ij} + \vec{F}^{\mathrm{D}}_{ij}
  + \vec{F}^{\mathrm{R}}_{ij}\,,
\end{equation}
where the conservative $\vec{F}^{\mathrm{C}}_{ij}$, dissipative
$\vec{F}^{\mathrm{D}}_{ij}$ and random $\vec{F}^{\mathrm{R}}_{ij}$
contributions are of the following form:
\begin{equation}\label{FC}
  \vec{F}^{\mathrm{C}}_{ij} =
     \left\{
     \begin{array}{ll}
        a(1-r_{ij})\hat{r}_{ij}, & r_{ij}<1,\\
        0,                       & r_{ij}\geqslant 1,
     \end{array}
     \right.
\end{equation}
\begin{equation}\label{FDR}
  \vec{F}^{\mathrm{D}}_{ij} =
  -\gamma w^{\mathrm{D}}(r_{ij})(\hat{r}_{ij}\cdot\vec{v}_{ij})\hat{r}_{ij},\qquad
  \vec{F}^{\mathrm{R}}_{ij} = \sigma w^{\mathrm{R}}(r_{ij})\theta_{ij}\Delta t^{-1/2}\hat{r}_{ij}.
\end{equation}
Here $\vec{v}_{ij}=\vec{v}_i-\vec{v}_j$, $\vec{v}_i$ and $\vec{v}_j$
being the velocities of the beads, $\theta_{ij}$ is Gaussian random
variable: $\langle \theta_{ij}(t) \rangle=0$, $\langle \theta_{ij}(t)
\theta_{kl}(t')\rangle=(\delta_{ik}\delta_{il} +
\delta_{il}\delta_{jk})\delta(t-t')$. According to Espa\~{n}ol and
Warren \cite{EspWarr}, the dissipative and random force amplitudes are
interrelated, $w^{\mathrm{D}}(r_{ij})=(w^{\mathrm{R}}(r_{ij}))^2$ and
$\sigma^2=2\gamma$, to satisfy the detailed balance. The frequently
used analytical form is:
\begin{equation}
w^{\mathrm{D}}(r_{ij})=(w^{\mathrm{R}}(r_{ij}))^2=
  \left\{
  \begin{array}{ll}
     (1-r_{ij})^2, & r_{ij}<1,\\
     0,            & r_{ij}\geqslant 1.
  \end{array}
  \right.
\end{equation}
We choose the value $\gamma=6.75$. Parameter $a$ in the conservative
force $\vec{F}^{\mathrm{C}}_{ij}$ defines maximum repulsion between
two beads which occurs at complete overlap $r_{ij}=0$. Three different
$a$ parameters can be distinguished, $a_{\mathrm{pp}}$ for the
polymer-polymer, $a_{\mathrm{ss}}$ for the solvent-solvent and
$a_{\mathrm{ps}}$ for the polymer-solvent interactions and all these
can be varied. As was shown in \cite{GrWarr}, these parameters can be
related to the Flory-Huggins parameter $\chi$. To integrate the
equations of motion, we used the modified velocity--Verlet algorithm
by Groot and Warren \cite{GrWarr}. Let us note that the structural
properties of the chain in a solvent are defined solely by the
conservative forces (\ref{FC}), while the dissipative and random
forces (\ref{FDR}) govern the phase space trajectory.

In our DPD simulations we considered the following set of chain
lengths: $N=5$, $7$, $10$, $14$, $20$, $28$, $40$, $56$ and $80$ beads
in four types of solvent:
\begin{equation}\label{solvents}
\begin{array}{lll}
\mbox{athermal solvent 1,} & a_{\mathrm{pp}}=a_{\mathrm{ss}}=25, & a_{\mathrm{ps}}=25,\\
\mbox{athermal solvent 2,} & a_{\mathrm{pp}}=a_{\mathrm{ss}}=33, & a_{\mathrm{ps}}=33,\\
\mbox{good solvent,}       & a_{\mathrm{pp}}=a_{\mathrm{ss}}=25, & a_{\mathrm{ps}}=20,\\
\mbox{very good solvent,}  & a_{\mathrm{pp}}=a_{\mathrm{ss}}=25, & a_{\mathrm{ps}}=10.
\end{array}
\end{equation}
The terms ``good'' and ``very good'' are indicative rather than exact,
since only repulsive forces are taken into account in DPD. Hence,
``good solvent'' means less repulsion for the polymer-solvent
interaction than that for the polymer-polymer one. The last two cases
in equations~(\ref{solvents}) correspond to the values $\xi=-0.2$ and
$\xi=-0.6$, respectively, in notation of Kong et al. \cite{Kong}.

All the simulations were performed in $NVT$ ensemble. At first, the
initial chain conformation was generated as a RW. Then, it was
immersed into a relatively large box filled by solvent and short
preliminary runs were performed to estimate the approximate
$R_{\mathrm{g}}$ for the polymer.  Thereafter, the final run was
performed using the cubic box of the linear size $L\approx
5R_{\mathrm{g}}$. This size is considered to be sufficiently large to
eliminate the effects of PBC \cite{Jiang,PierRyk}. The number of
solvent beads $N_{\mathrm{s}}$ was chosen to keep the total reduced
density of the solution equal to $\rho^*=(N+N_{\mathrm{s}})/V=3$. The
time step was chosen to be $\Delta t=0.04$ in reduced units. We
performed from $3\cdot 10^{6}$ (for $N=5$) up to $15\cdot 10^{6}$ (for
$N=80$) DPD steps for the final run.  We did not concentrate on the
relaxation times in detail, but found the behaviour similar to Jiang
et al. \cite{Jiang}, $\tau\approx 0.13\,N^{1.81}$ who used similar DPD
parameters for athermal solvent.  This brings us to the following
estimates for the simulation runs: about $5\cdot 10^4\,\tau$ for $N=5$
and $1.7\cdot 10^3\,\tau$ for $N=80$ with the estimates for
intermediate values of $N$ falling in between. The chain conformations
were saved after performing every $1000$ DPD steps and last $3/4$ of
these were used for the subsequent analysis. As far as all the
simulations were performed using PBC, the unwrapping of the chain at
each time step was performed based on the connectivity information.

\section{Results}\label{sec4}

We concentrated our analysis on the three following metric
characteristics of the linear polymer. The instantaneous squared
end-to-end distance at a given time $t$ is defined as:
\begin{equation}\label{R1N}
R_{1N}^2(t) = (\vec{r}_1-\vec{r}_N)^2,
\end{equation}
where $\vec{r}_1$ and $\vec{r}_N$ are the radius-vectors for the first
and last bead of the chain, respectively. The squared radius of
gyration at time $t$ is defined as:
\begin{equation}\label{Rg}
R_{\mathrm{g}}^2(t) = \frac{1}{N}\sum_{i=1}^{N}(\vec{r}_i-\vec{R}_{\rm COM})^2,
\end{equation}
where $\vec{R}_{\rm COM}$ is the radius-vector of the polymer center of
mass. The inverse hydrodynamic radius is defined as follows:
\begin{equation}\label{Rh}
\frac{1}{R_{\mathrm{h}}(t)} = \frac{1}{N^2}\sum_{i\neq j}\frac{1}{r_{ij}},
\end{equation}
where the sum runs over all different bead pairs $i,j$. All three
properties are time averaged (denoted hereafter as
$\langle\cdots\rangle$) and then the first two can be square rooted
while the third one is inverted to provide the estimates for $R_{1N}$,
$R_{\mathrm{g}}$ and $R_{\mathrm{h}}$, respectively.

One should consider $R_{\mathrm{g}}$ and $R_{\mathrm{h}}$ as more
useful characteristics as compared to $R_{1N}$. First of all, $R_{1N}$
is meaningful for the linear polymer or for a certain subchain of the
branched molecule (e.g. backbone, side chain, star-polymer arm, etc.),
while both $R_{\mathrm{g}}$ and $R_{\mathrm{h}}$ are universally
defined. Apart from that, both these properties are directly
measurable via small-angle scattering experiments and via small-angle
dynamic light scattering experiments, respectively (for more
discussion on this, see~\cite{Dunweg}).

\subsection{Linear fit and first correction-to-scaling fit}

As was already mentioned above, a number of universal scaling laws
have been observed for the polymer properties \cite{deGennes79} in the
limit of infinite polymer length, e.g.  equations~(\ref{1}). To
reproduce these properties, one needs to consider rather long chains
in the simulations \cite{LiMadr}. Due to the polymer coil
self-similarity, one would expect that such a coarse-grained approach
as DPD should be capable of reproducing correct scaling laws for much
shorter chain lengths. For the moderate polymer lengths, usually the
scaling is written in terms of a number of bonds, $N-1$, instead of
the number of monomers $N$. Also, as is known, the
correction-to-scaling terms are important for moderate system sizes,
as is indicated by the studies of the phase transitions in spin
lattice models \cite{Ballesteros99,Ballesteros98,Folk99}.  Taking into
account the first correction-to-scaling term, equation~(\ref{2}), one
has for the $\langle R_{1N}^2\rangle$:
\begin{equation}
\langle R_{1N}^2 \rangle = A (N-1)^{2\nu}\left(1 + \frac{B}{(N-1)^{\Delta}}+\cdots\right).
\end{equation}
We shall work with  square rooted values
\begin{equation}\label{R1Nscaling}
\Rn = \sqrt{\langle R_{1N}^2 \rangle}\approx A^{1/2}(N-1)^{\nu}\left(1 +
\frac{1}{2}\frac{B}{(N-1)^{\Delta}}+\cdots\right),
\end{equation}
and keep the linear term only in the correction-to-scaling expansion above. Taking a logarithm of
both sides of equations~(\ref{R1Nscaling}), we obtain
\begin{eqnarray}
\ln \Rn&=&\frac{1}{2}\ln A+\nu\ln(N-1)+
          \ln\left(1 + \frac{1}{2}\frac{B}{(N-1)^{\Delta}}+\cdots\right)\nonumber\\
       &=&A'+\nu\ln(N-1)+ \ln\left(1 + \frac{B'}{(N-1)^{\Delta}}+\cdots\right),
       \label{FCSwork}
\end{eqnarray}
where $A'$ and $B'$ are self-explanatory. The same formulae can also
be applied for $\Rg=\sqrt{\langle R_{\mathrm{g}}^2 \rangle}$ and
$\Rh=1/\langle 1/R_{\mathrm{h}} \rangle$ (in the latter case one has
$B'\approx -B$). As a result, the same fit (\ref{FCSwork}) can be used
for all three quantities under interest, $\Rn$, $\Rg$ and $\Rh$. In
the case of linear fit the correction-to-scaling amplitude $B'$ is
assumed to be zero and one recovers the linear fit form.  To perform
the fits we used the least-squares routine {\it mrqmin} from the
Numerical recipes book \cite{NumRec}. The accuracy of the fit was
monitored by the cumulative squared deviation per one data point,
\begin{equation}
\chi^2=\frac{1}{n_{\rm data}}\sum_{k=1}^{n_{\rm data}}
       \left[\Rn^{[k]}-F(N^{[k]})\right]^2,
\end{equation}
where ($N^{[k]}$,$\Rn^{[k]}$) is the $k$-th out of $n_{\rm data}$ data
points and $F(N)$ is the fitting function.

It is generally known that very good statistics for each data point
are required to ensure robust results for the scaling exponent
$\nu$. Otherwise the latter depends essentially on the selection of
data points being used for the fit. This was also found to be the case
in the course of the present study. Hence, relatively lengthly
simulations are performed to improve statistics. To check the
consistency of the results, we performed a series of fits using a
gradually increasing range of intervals in the chain length $N$,
e.g. $N=5\div28, N=5\div40, \cdots N=5\div80$, which are presented
below.

\begin{figure}[h]
\centering
\includegraphics[width=7cm]{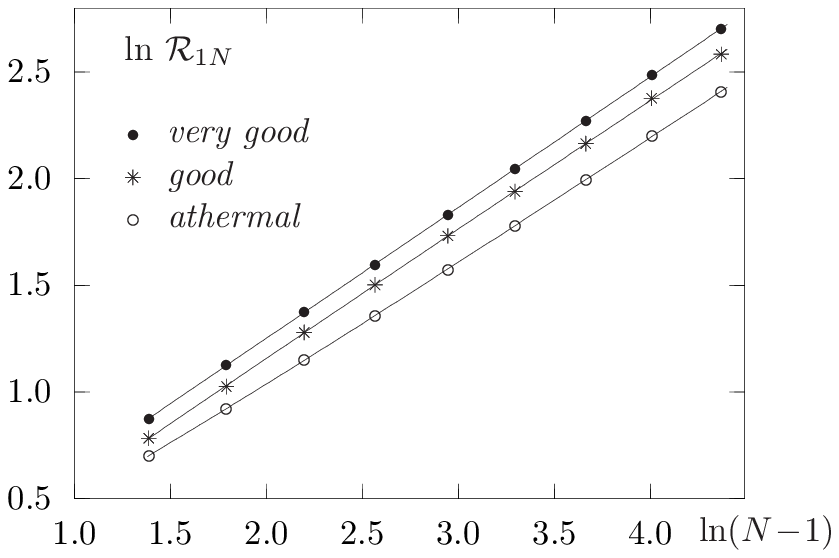}~~
\includegraphics[width=7cm]{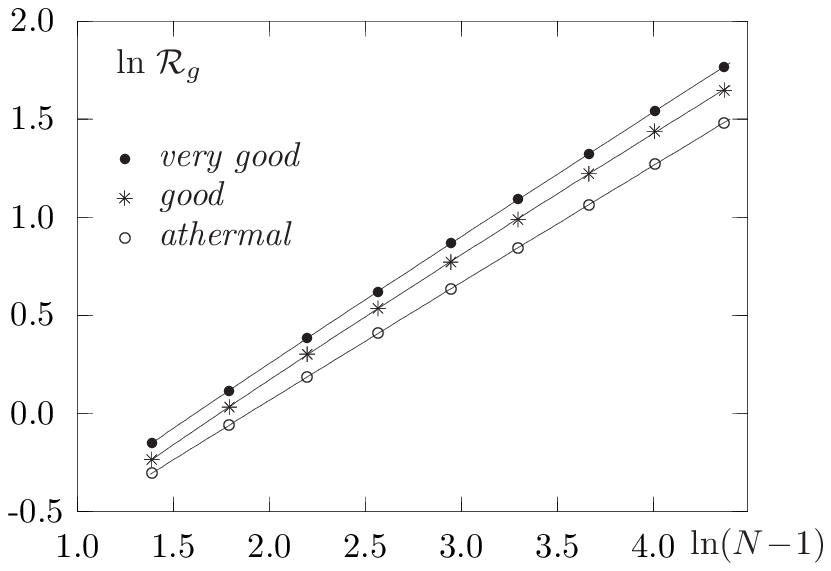}
\caption{\label{fig_r1n_rg}Data points for $\Rn$ (on the left) and
  $\Rg$ (on the right) and their best fits using the form
  \protect(\ref{FCSwork}) obtained from the DPD simulations on a set of
  chain lengths $N=5\div80$ at different solvent conditions, see
  notations on both plots.}
\end{figure}

As was anticipated, both cases of ``athermal 1'' and ``athermal 2''
solvents were found to show the same averaged properties within the
accuracy of the simulations. Therefore, the data for both cases have
been averaged at each $N$ and marked as ``athermal'' further in the
text. At first glance, the data points for both $\Rn$ and $\Rg$ are
well fitted by the line in log-log scale (see,
figure~\ref{fig_r1n_rg}), but, in fact, the best fits were achieved
using the first correction-to-scaling form (\ref{FCSwork}). We should
remark here that we were unable to achieve stable results for both
$\Rn$ and $\Rg$ using the four-parameter fits, in which all $A'$,
$\nu$, $B'$ and $\Delta$ in (\ref{FCSwork}) are fitted
simultaneously. Instead, we were forced to fix the
correction-to-scaling exponent $\Delta$ at certain value and to
perform the fit over three remaining parameters. We used two choices
for $\Delta$, namely the best theoretical value $\Delta=0.478$
\cite{Guida98} and $\Delta=1$.

The results presented in table~\ref{tab_R1N} indicate that the
correction-to-scaling term for $\Rn$ is essential in the case of
athermal solvent only. In the table, we also show the linear fit
performed over the largest chain lengths, $N=20\div80$, for the sake
of comparison. One can see that the latter leads to almost the same
result $\nu=0.586$ as the correction-to-scaling fit $\nu=0.591$, which
is indicative of the crossover quickly decreasing at $N>20$. In the
cases of good and very good solvents very small amplitudes $B'$ are
found and, as the result, linear fits are almost equally good (see,
table~\ref{tab_R1N}). For $\Rg$ the situation is exactly reversed. The
correction to scaling is important for the cases of good and very good
solvents (see, table~\ref{tab_Rg}). We cannot suggest some definite
explanation for this effect, except the conclusion that preference
should be given to the general form (\ref{FCSwork}), rather than to
the linear fit, otherwise the exponent $\nu$ could be essentially
under- or overestimated (see, tables~\ref{tab_R1N}, \ref{tab_Rg}).
\begin{table}[!h]
\caption{\label{tab_R1N} Linear fit and first correction-to-scaling
  fits to the form \protect(\ref{FCSwork}) for $\Rn$ with fixed
  exponent $\Delta=0.478$ and $\Delta=1$ when using various sets of
  data points. The best fit for the interval $N=5\div80$ is
  underlined.}\vspace{1ex}
\begin{center}
\begin{tabular}{|l|l|l|l|}
\hline
fitting & linear fit & ~~~fit with $\Delta=0.478$ & ~~~fit with $\Delta=1$\\
range, $N$ &~~~~$\nu$~~~~~~$\chi^2$ & ~~~~$\nu$~~~~~~$[B']$~~~~~~$\chi^2$ &
~~~~$\nu$~~~~~~$[B']$~~~~~~$\chi^2$\\
\hline
\multicolumn{4}{|c|}{athermal solvent}\\
\hline ~~5~$\div$~28~ & ~0.564~~\error{1.7e--5}~ & ~0.617~~[0.370]~~\error{9.4e--7}~ &
                                      ~0.588~~[0.219]~~\error{1.3e--6}~ \\
~~5~$\div$~40~ & ~0.569~~\error{3.1e--5}~ & ~0.621~~[0.399]~~\error{9.0e--7}~ &
                                      ~0.592~~[0.252]~~\error{1.8e--6}~ \\
~~5~$\div$~56~ & ~0.573~~\error{5.5e--5}~ & ~0.627~~[0.447]~~\error{1.3e--6}~ &
                                      ~0.596~~[0.295]~~\error{3.6e--6}~ \\
~~5~$\div$~80~ & ~0.574~~\error{5.2e--5}~ & ~0.611~~[0.318]~~\error{8.2e--6}~ &
                                \best{~0.591}~~[\best{0.248}]~~\best{\error{6.9e--6}~}\\
~20~$\div$~80~ & ~0.586~~\error{8.6e--6}~ &&\\
\hline
\multicolumn{4}{|c|}{good solvent}\\
\hline ~~5~$\div$~28~ & ~0.607~~\error{8.6e--6}~ & ~0.591~~[--0.099]~~\error{6.4e--6}~ &
                                      ~0.600~~[--0.066]~~\error{6.9e--6}~ \\
~~5~$\div$~40~ & ~0.607~~\error{7.7e--6}~ & ~0.595~~[--0.076]~~\error{5.7e--6}~ &
                                      ~0.601~~[--0.056]~~\error{6.0e--6}~ \\
~~5~$\div$~56~ & ~0.607~~\error{7.3e--6}~ & ~0.604~~[--0.026]~~\error{7.0e--6}~ &
                                      ~0.605~~[--0.023]~~\error{7.0e--6}~ \\
~~5~$\div$~80~ & ~0.605~~\error{1.7e--5}~ &
\best{~0.594}~~[\best{--0.087}]~~\best{\error{1.1e--5}~} &
                                      ~0.600~~[--0.073]~~\error{1.2e--5}~ \\
~20~$\div$~80~ & ~0.601~~\error{1.5e--5}~ &&\\
\hline
\multicolumn{4}{|c|}{very good solvent}\\
\hline ~~~5~$\div$~28~ & ~0.614~~\error{6.3e--7}~ & ~0.610~~[--0.021]~~\error{5.4e--7}~ &
                                       ~0.612~~[--0.019]~~\error{5.0e--7}~ \\
~~~5~$\div$~40~ & ~0.613~~\error{1.7e--6}~ & ~0.605~~[--0.052]~~\error{8.3e--7}~ &
                                       ~0.609~~[--0.041]~~\error{8.4e--7}~ \\
~~~5~$\div$~56~ & ~0.613~~\error{2.1e--6}~ & ~0.612~~[--0.009]~~\error{2.1e--6}~ &
                                       ~0.612~~[--0.011]~~\error{2.1e--6}~ \\
~~~5~$\div$~80~ & ~0.613~~\error{2.0e--6}~ & ~0.611~~[--0.012]~~\error{1.9e--6}~ &
                                 \best{~0.612}~~[\best{--0.012}]~~\best{\error{1.8e--6}~} \\
~20~$\div$~80~  & ~0.613~~\error{2.8e--6}~ &&\\
\hline
\end{tabular}
\end{center}
\end{table}
\begin{table}[ht]
\caption{\label{tab_Rg}Linear fit and first correction-to-scaling fits
  to the form \protect(\ref{FCSwork}) for $\Rg$ with fixed exponent
  $\Delta=0.478$ and $\Delta=1$ when using various sets of data
  points. The best fit for the interval $N=5\div80$ is underlined.}
  \vspace{1ex}
\begin{center}
\begin{tabular}{|l|l|l|l|}
\hline
fitting & linear fit & ~~~fit with $\Delta=0.478$ & ~~~fit with $\Delta=1$\\
range, $N$ &~~~~$\nu$~~~~~~$\chi^2$ & ~~~~$\nu$~~~~~~$[B']$~~~~~~$\chi^2$ &
~~~~$\nu$~~~~~~$[B']$~~~~~~$\chi^2$\\
\hline
\multicolumn{4}{|c|}{athermal solvent}\\
\hline ~~5~$\div$~28~ & ~0.601~~\error{2.7e--6}~ & ~0.584~~[--0.106]~~\error{2.0e--7}~ &
                                      ~0.592~~[--0.080]~~\error{2.5e--7}~ \\
~~5~$\div$~40~ & ~0.600~~\error{3.1e--6}~ & ~0.587~~[--0.090]~~\error{3.0e--7}~ &
                                      ~0.593~~[--0.073]~~\error{2.6e--7}~ \\
~~5~$\div$~56~ & ~0.600~~\error{2.8e--6}~ & ~0.593~~[--0.052]~~\error{1.5e--6}~ &
                                      ~0.597~~[--0.047]~~\error{1.3e--6}~ \\
~~5~$\div$~80~ & ~0.599~~\error{8.0e--6}~ &
\best{~0.587}~~[\best{--0.087}]~~\best{\error{2.9e--6}~} &
                                      ~0.593~~[--0.077]~~\error{3.1e--6}~ \\
~20~$\div$~80~ & ~0.596~~\error{6.0e--6}~ &&\\
\hline
\multicolumn{4}{|c|}{good solvent}\\
\hline ~~5~$\div$~28~ & ~0.641~~\error{4.9e--5}~ & ~0.573~~[--0.368]~~\error{2.7e--6}~ &
                                      ~0.605~~[--0.318]~~\error{4.2e--6}~ \\
~~5~$\div$~40~ & ~0.638~~\error{5.3e--5}~ & ~0.586~~[--0.314]~~\error{5.4e--6}~ &
                                      ~0.610~~[--0.283]~~\error{5.0e--6}~ \\
~~5~$\div$~56~ & ~0.636~~\error{5.1e--5}~ & ~0.597~~[--0.260]~~\error{9.4e--6}~ &
                                      ~0.615~~[--0.244]~~\error{7.3e--6}~ \\
~~5~$\div$~80~ & ~0.631~~\error{9.7e--5}~ &
\best{~0.588}~~[\best{--0.306}]~~\best{\error{1.4e--5}~} &
                                      ~0.609~~[--0.299]~~\error{1.4e--5}~ \\
~20~$\div$~80~ & ~0.618~~\error{2.8e--5}~ &&\\
\hline
\multicolumn{4}{|c|}{very good solvent}\\
\hline ~~5~$\div$~28~ & ~0.651~~\error{6.9e--6}~ & ~0.626~~[--0.144]~~\error{1.9e--6}~ &
                                      ~0.639~~[--0.109]~~\error{2.2e--6}~ \\
~~5~$\div$~40~ & ~0.647~~\error{1.6e--5}~ & ~0.618~~[--0.185]~~\error{2.6e--6}~ &
                                      ~0.633~~[--0.151]~~\error{3.6e--6}~ \\
~~5~$\div$~56~ & ~0.645~~\error{2.4e--5}~ & ~0.616~~[--0.194]~~\error{2.4e--6}~ &
                                      ~0.631~~[--0.168]~~\error{3.7e--6}~ \\
~~5~$\div$~80~ & ~0.642~~\error{3.4e--5}~ &
\best{~0.615}~~[\best{--0.201}]~~\best{\error{2.2e--6}~} &
                                      ~0.629~~[--0.184]~~\error{3.9e--6}~ \\
~20~$\div$~80~ & ~0.632~~\error{2.3e--6}~ &&\\
\hline
\end{tabular}
\end{center}
\end{table}
\begin{table}[!h]
\caption{\label{tab_Rh}Linear fit and first correction-to-scaling fits
  to the form \protect(\ref{FCSwork}) for $\Rh$ with fixed exponent
  $\Delta=0.478$ and from the four-parameter fits when using various
  sets of data points. The best fit for the interval $N=5\div80$ is
  underlined.}
  \vspace{1ex}
\begin{center}
\begin{tabular}{|l|l|l|l|l|}
\hline
fitting & ~~~fit with $\Delta=0.478$ & ~~~four-parameter fit\\
range, $N$ &~~~~$\nu$~~~~~~$[B']$~~~~~~$\chi^2$ &
~~~~$\nu$~~~~~~$\Delta$~~~~~~$[B']$~~~~~~$\chi^2$\\
\hline
\multicolumn{3}{|c|}{athermal solvent}\\
\hline
~~5~$\div$~28~ & 0.596~~[3.008]~~\error{4.9e--5}~ & ~0.497~~0.836~~[1.324]~~\error{1.9e--7}~ \\
~~5~$\div$~40~ & 0.611~~[3.275]~~\error{6.9e--5}~ & ~0.534~~0.732~~[1.683]~~\error{4.7e--7}~ \\
~~5~$\div$~56~ & 0.625~~[3.555]~~\error{9.5e--5}~ & ~0.572~~0.666~~[2.195]~~\error{1.0e--6}~ \\
~~5~$\div$~80~ & 0.635~~[3.786]~~\error{1.0e--4}~ &
                    \best{~0.552}~~\best{0.699}~~[\best{1.912}]~~\best{\error{1.2e--6}~}\\
&&\\
\hline
\multicolumn{3}{|c|}{good solvent}\\
\hline
~~5~$\div$~28~ & 0.623~~[2.953]~~\error{2.0e--5}~ & ~0.476~~1.017~~[0.925]~~\error{4.8e--7}~ \\
~~5~$\div$~40~ & 0.639~~[3.274]~~\error{3.1e--5}~ & ~0.563~~0.654~~[1.636]~~\error{2.3e--6}~ \\
~~5~$\div$~56~ & 0.654~~[3.592]~~\error{5.1e--5}~ & ~0.641~~0.577~~[2.993]~~\error{3.0e--6}~ \\
~~5~$\div$~80~ & 0.658~~[3.710]~~\error{4.7e--5}~ &
                    \best{~0.565}~~\best{0.664}~~[\best{1.663}]~~\best{\error{5.5e--6}~}\\
&&\\
\hline
\multicolumn{3}{|c|}{very good solvent}\\
\hline
~~5~$\div$~28~ & 0.625~~[3.052]~~\error{5.9e--5}~ & ~0.652~~0.643~~[3.159]~~\error{1.5e--7}~ \\
~~5~$\div$~40~ & 0.639~~[3.284]~~\error{7.8e--5}~ & ~0.580~~0.703~~[1.925]~~\error{6.1e--7}~ \\
~~5~$\div$~56~ & 0.651~~[3.533]~~\error{9.7e--5}~ & ~0.583~~0.698~~[1.963]~~\error{5.4e--7}~ \\
~~5~$\div$~80~ & 0.662~~[3.779]~~\error{1.1e--4}~ &
                    \best{~0.577}~~\best{0.710}~~[\best{1.885}]~~\best{\error{5.1e--7}~}\\
&&\\
\hline
\end{tabular}
\end{center}
\end{table}

\begin{figure}[h]
\centering
\includegraphics[width=7cm]{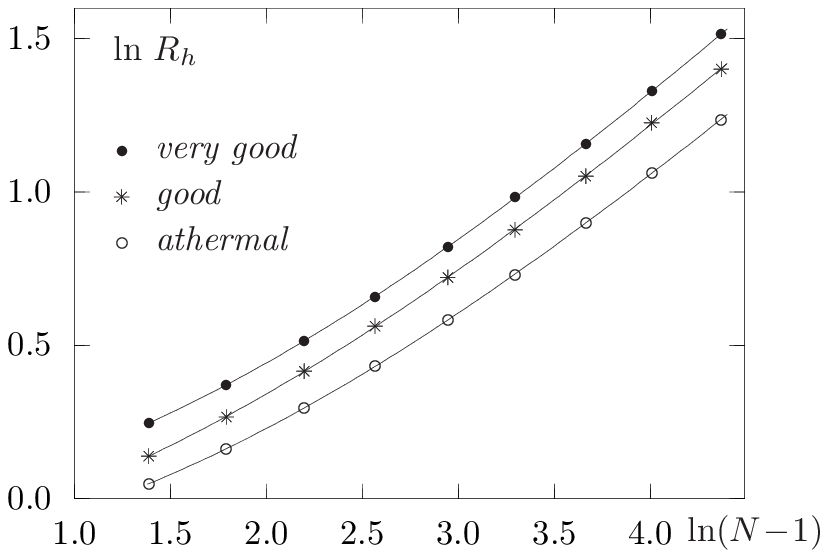}
\caption{\label{fig_rh}Data points for $\Rh$ and their best fits using the
  form \protect(\ref{FCSwork}) obtained from the DPD simulations on a
  set of chain lengths $N=5\div80$ at different solvent conditions, see
  notations on the plot.}
\end{figure}

The four-parameter fit, mentioned above, does not work for either
$\Rn$ or $\Rg$ but is supposed to perform better for $\Rh$, where
essentially larger correction to scaling is to be expected
\cite{Dunweg}. Indeed, this is found to be the case in our
simulations. First of all, this can be noticed from the arrangement of
the data points, see figure~\ref{fig_rh}. The results based on
numerical fits are presented in table~\ref{tab_Rh}, where both the
cases of fixed $\Delta=0.478$ and of the four-parameter fits are
shown. The latter give essentially better results in terms of fitting
accuracy and provide independent estimates for both exponents,
$\nu\approx 0.56\div0.58$ and $\Delta\approx 0.65\div0.70$. As was
demonstrated by Dunweg et al. \cite{Dunweg} there are two comparable
correction-to-scaling terms for the $\Rh$, one is proportional to
$N^{-\Delta}$ and another, the so-called ``analytic'' correction, is
proportional to $N^{-(1-\nu)}$. Therefore, if one uses the form
(\ref{FCSwork}) with a single correction term, then a rather effective
exponent $\Delta$ will be obtained. This is how we should interpret
the value $\Delta$ obtained in our study.

\subsection{The fits based on the distribution of $\Rg$ values}

The averaging of the metric properties of interest,
e.g. $\Rg=\sqrt{\langle R_{\mathrm{g}}^2 \rangle}$, throughout this
study was performed as the arithmetic mean over discrete set of time
points. One can also be interested in the distribution
$\bar{p}(R_{\mathrm{g}})$ of the $R_{\mathrm{g}}$ values themselves
and, for instance, in examination of the scaling law for the most
probable values with the chain length $N$. The following analytical
form has been postulated by Lhuillier in 3d \cite{Lhuill}
\begin{equation}\label{Lhuillier}
  \bar{p}(R_{\mathrm{g}}) \sim \exp\left[-A_1(N^\nu/R_{\mathrm{g}})^{3\alpha}
                        -A_2(R_{\mathrm{g}}/N^\nu)^\delta\right],
\end{equation}
where $\alpha=1/(3\nu-1)$, $\delta=1/(1-\nu)$ and $A_1$, $A_2$ are
constants. Subsequently, this has been verified in a number of lattice
MC simulations \cite{Victor,Bishop}.  One can rewrite this
distribution in terms of $R_{\mathrm{g}}^2$, using the expressions for
$\alpha$ and $\delta$ and incorporating the $N$-dependent factors into
the constants, and the result reads:
\begin{equation}\label{Lhuilfit}
  p(R_{\mathrm{g}}^2) = C \exp\left[-\frac{a_1}{[R_{\mathrm{g}}^2]^{\frac{3}{2(3\nu-1)}}}
                               -a_2[R_{\mathrm{g}}^2]^{\frac{1}{2(1-\nu)}}\right].
\end{equation}
This distribution has a maximum at
\begin{equation}\label{Rgmax}
  [R_{\mathrm{g}}^2]^{\max}=\left[\frac{3a_1(1-\nu)}{a_2(3\nu-1)}\right]^
             {(1-\nu)(3\nu-1)}
\end{equation}
which can be square rooted to obtain an estimate for the most probable
value of the radius of gyration
$\Rg^{\max}=\sqrt{[R_{\mathrm{g}}^2]^{\max}}$. We prefer to work with
the distribution of $R_{\mathrm{g}}^2$ (\ref{Lhuilfit}) and then to
square root the maximum to match the averaging procedure for simple
averaging, $\Rg=\sqrt{\langle R_{\mathrm{g}}^2\rangle}$.

From the point of view of numerical fitting, the form (\ref{Lhuilfit})
has four parameters: one exponent $\nu$ and three coefficients, $a_1$,
$a_2$ and $C$. Therefore, one can have a first estimate for the
exponent $\nu$ right from the fit of the data to the form
(\ref{Lhuilfit}).  The second estimate for $\nu$ can be obtained from
the anticipated scaling of $\Rg^{\max}$ according to the form
(\ref{FCSwork}).

\begin{figure}[!h]
\centering
\includegraphics[width=9cm]{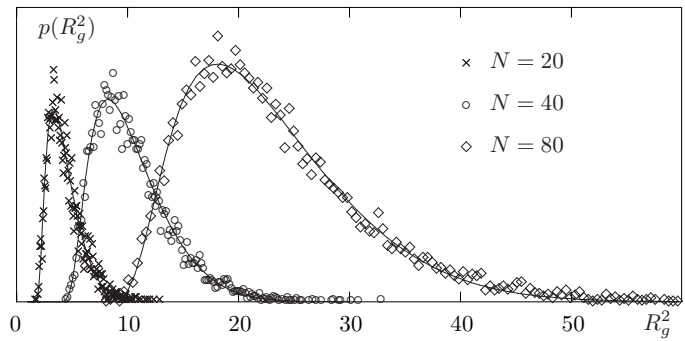}
\caption{\label{Lhuilplot}Mapping of the distribution of radius of
  gyration values $p(R_{\mathrm{g}}^2)$ obtained from the DPD
  simulations in athermal solvent, onto Lhuillier form
  \protect(\ref{Lhuilfit}), a few indicated chain lengths are shown
  only for the sake of clarity.}
\end{figure}
\begin{figure}[h]
\centering
\includegraphics[width=7cm]{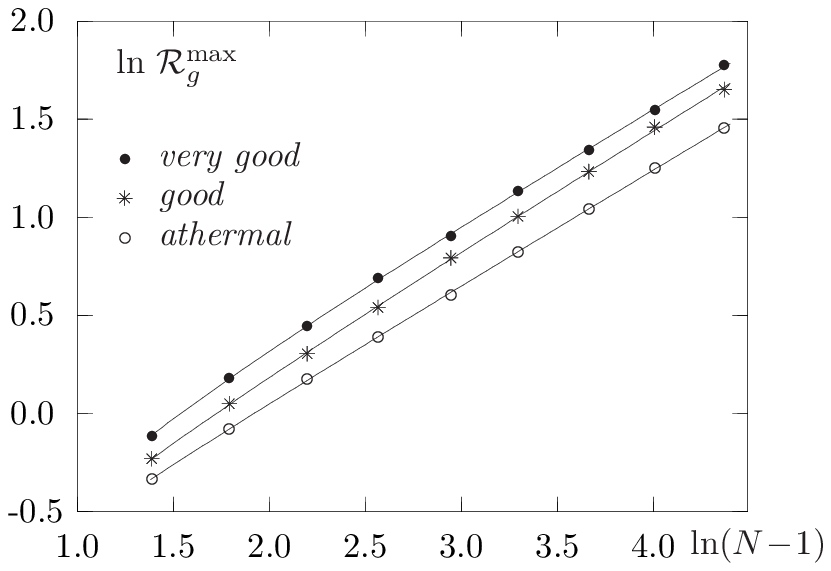}
\caption{\label{fig_rg_max}Data points for $\Rg^{\max}$ and their best
  fits using the form \protect(\ref{FCSwork}) obtained from the DPD
  simulations on a set of chain lengths $N=5\div80$ at different solvent
  conditions, see notations on the plot.}
\end{figure}

\looseness=-1The distributions $p(R_{\mathrm{g}}^2)$ were built in a
form of histograms and were found to map very well on the Lhuillier
form (\ref{Lhuilfit}), see samples in figure~\ref{Lhuilplot} obtained
for the case of athermal solvent. These fits, as was mentioned above,
provide both the estimate for $\Rg^{\max}$ and an independent estimate
for the exponent $\nu$. The latter was found to be scattered in the
interval $\nu=0.53\div0.58$ depending on $N$ and solvent type. We do
not expect high accuracy from such a fitting, as far as dependence on
$\nu$ in equations~(\ref{Lhuilfit}) is rather complex. Nevertheless,
the values of $\nu$ found are very reasonable and indicate a
self-consistency of this numerical approach. The next step is to use
the maxima positions $\Rg^{\max}$ of the distributions
(\ref{Lhuilfit}), found via equations~(\ref{Rgmax}), and to find a fit
to the scaling law (\ref{FCSwork}) similarly to the analysis of $\Rg$
data in the previous subsection. The data points alongside with the
best fits are shown in figure~\ref{fig_rg_max} and the results for the
exponent $\nu$, correction-to-scaling amplitude $B'$ and fitting error
$\chi^2$ are presented in table~\ref{tab_rg_max}. We would like to
stress that, contrary to the results given in table~\ref{tab_Rg}, no
dependence on solvent quality is observed in table~\ref{tab_rg_max}
and the same value $\nu=0.582-0.585$ is found essentially in all three
cases. This might indicate that the effect of small drift of $\nu$
observed in a very good solvent in table~\ref{tab_Rg} could be an
artefact of the numerical method used.
\begin{table}[ht]
\caption{\label{tab_rg_max}Linear fit and first correction-to-scaling
  fits to the form \protect(\ref{FCSwork}) for $R_{\mathrm{g}}^{\max}$ with fixed
  exponent $\Delta=0.478$ and $\Delta=1$ when using various sets of
  data points. The best fit for the interval $N=5\div80$ is underlined.}
  \vspace{1ex}
\begin{center}
\begin{tabular}{|l|l|l|l|}
\hline
fitting & linear fit & ~~~fit with $\Delta=0.478$ & ~~~fit with $\Delta=1$\\
range, $N$ &~~~~$\nu$~~~~~~$\chi^2$ &
~~~~$\nu$~~~~~~$[B']$~~~~~~$\chi^2$ &
~~~~$\nu$~~~~~~$[B']$~~~~~~$\chi^2$\\
\hline
\multicolumn{4}{|c|}{athermal solvent}\\
\hline ~~5~$\div$~28~ & ~0.602~~\error{6.4e--5}~ & ~0.539~~[--0.345]~~\error{2.6e--5}~ &
                                      ~0.568~~[--0.306]~~\error{2.4e--5}~ \\
~~5~$\div$~40~ & ~0.601~~\error{5.6e--5}~ & ~0.562~~[--0.242]~~\error{3.2e--5}~ &
                                      ~0.579~~[--0.222]~~\error{2.8e--5}~ \\
~~5~$\div$~56~ & ~0.601~~\error{4.9e--5}~ & ~0.576~~[--0.169]~~\error{3.4e--5}~ &
                                      ~0.587~~[--0.165]~~\error{3.0e--5}~ \\
~~5~$\div$~80~ & ~0.598~~\error{5.6e--5}~ & ~0.574~~[--0.183]~~\error{3.1e--5}~ &
                                      \best{~0.585}~~[\best{--0.181}]~~\best{\error{2.7e--5}}~ \\
~20~$\div$~80~ & ~0.598~~\error{2.3e--5}~ &&\\
\hline
\multicolumn{4}{|c|}{good solvent}\\
\hline ~~5~$\div$~28~ & ~0.646~~\error{4.6e--5}~ & ~0.594~~[--0.291]~~\error{2.1e--5}~ &
                                      ~0.618~~[--0.247]~~\error{2.0e--5}~ \\
~~5~$\div$~40~ & ~0.641~~\error{6.6e--5}~ & ~0.588~~[--0.316]~~\error{1.8e--5}~ &
                                      ~0.613~~[--0.283]~~\error{1.9e--5}~ \\
~~5~$\div$~56~ & ~0.640~~\error{5.9e--5}~ & ~0.606~~[--0.231]~~\error{2.7e--5}~ &
                                      ~0.622~~[--0.217]~~\error{2.5e--5}~ \\
~~5~$\div$~80~ & ~0.633~~\error{1.8e--4}~ &
\best{~0.582}~~[\best{--0.352}]~~\best{\error{6.2e--5}}~ &
                                      ~0.607~~[--0.344]~~\error{6.9e--5}~ \\
~20~$\div$~80~ & ~0.609~~\error{9.4e--5}~ &&\\
\hline
\multicolumn{4}{|c|}{very good solvent}\\
\hline ~~5~$\div$~28~ & ~0.646~~\error{2.6e--4}~ & ~0.517~~[--0.626]~~\error{6.0e--5}~ &
                                      ~0.571~~[--0.634]~~\error{5.5e--5}~ \\
~~5~$\div$~40~ & ~0.636~~\error{3.4e--4}~ & ~0.522~~[--0.610]~~\error{5.2e--5}~ &
                                      ~0.570~~[--0.637]~~\error{4.7e--5}~ \\
~~5~$\div$~56~ & ~0.628~~\error{3.8e--4}~ & ~0.529~~[--0.582]~~\error{4.9e--5}~ &
                                      ~0.572~~[--0.624]~~\error{4.2e--5}~ \\
~~5~$\div$~80~ & ~0.625~~\error{3.7e--4}~ & ~0.549~~[--0.503]~~\error{7.8e--5}~ &
                                      \best{~0.582}~~[\best{--0.551}]~~\best{\error{5.9e--5}}~\\
~20~$\div$~80~ & ~0.607~~\error{3.9e--5}~ &&\\
\hline
\end{tabular}
\end{center}
\end{table}

\begin{figure}[h]
\centering
\includegraphics[width=8cm]{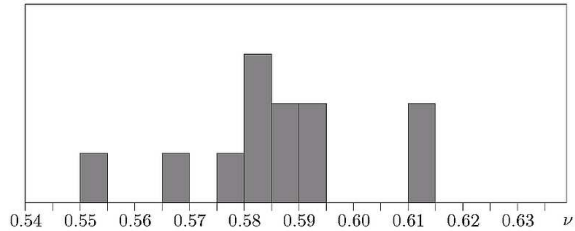}
\caption{\label{fig_nu_hist} A histogram of values of $\nu$ obtained
  for different characteristic radii of a polymer chain in different
  solvents (\protect\ref{solvents}). The data are collected according
  to best fits results (underlined in
  tables~\protect\ref{tab_R1N}--\protect\ref{tab_rg_max}). The value
  of $\nu$ is found predominantly in the interval $\nu=0.55\div0.61$
  and concentrate around the best theoretical estimate $\nu=0.5882\pm
  0.0011$ \cite{Guida98}.}
\end{figure}

The cumulative outcome of our analysis for the scaling exponent $\nu$
which governs the scaling law (\ref{2}) is presented in
figure~\ref{fig_nu_hist}. Here we show a histogram of the values of
$\nu$ obtained for $\Rn$, $\Rg$ and $\Rh$ of a polymer chain in
different solvents considered in the current study, namely, in
athermal, good and very good solvents. To make this histogram, the
best fits (underlined results in
tables~\ref{tab_R1N}--\ref{tab_rg_max}) have been used. One can see
from the histogram that the value of $\nu$ is scattered in an interval
$\nu=0.55\div0.61$ but in fact is found predominantly in much narrower
interval $\nu=0.58\div0.60$ centered around the best theoretical
estimate $\nu=0.5882\pm 0.0011$ \cite{Guida98}. This allows us to
conclude that within the accuracy of our analysis, the scaling law
(\ref{2}) holds for all polymer characteristics and in all solvents
considered in this study. Our conservative estimate for the scaling
exponent therefore is
\begin{equation}
\nu \approx 0.59 \pm 0.01.
\end{equation}

\section{Conclusions and outlook}\label{sec5}

In this study we performed DPD simulations of the polymer chain in a
solvent of various quality with the aim to reexamine how well the
scaling laws hold for various polymer characteristics. Chains of up to
$80$ beads were considered and simulation runs from $1600$ up to
$50000$ relaxation times were performed. Three metric properties were
considered, end-to-end distance, radius of gyration and hydrodynamic
radius. For the analysis we used the linear fits, first
correction-to-scaling fits and fitting the distribution of the radius
of gyration. As an outcome, within the accuracy of the simulations and
of the data analysis technique, the following conclusions can be
drawn:
\begin{itemize}
\item[(i)] All three metric properties obey the scaling law with very
  close values for $\nu$ found within the interval $0.55\div0.61$
  (depending on the method of analysis). Most values are concentrated
  around the average $\nu=0.59$ being very close to the theoretical
  estimate $\nu=0.5882\pm 0.0011$ \cite{Guida98}, see,
  figure~\ref{fig_nu_hist}.
\item[(ii)] Corrections to scaling are found to be important in all
  combinations except a few property/solvent ones and thus the
  conclusion (i) is valid only when the correction-to-scaling terms
  are taken into account.
\item[(iii)] No or very small (up to 4\%) drift of the scaling
  exponent $\nu$ is observed when changing the solvent quality
  (remaining, however, in a good solvent regime), but this effect
  depends on a numerical technique being used.
\end{itemize}

The method of analysis can be extended to the study of the scaling
laws in more complex molecular architectures, e.g. star-like polymers,
branched and hyperbranched molecules (including amphiphiles).

\section*{Acknowledgements}

We thank Prof. Myroslav Holovko for the invitation to contribute to
the Festschrift dedicated to Prof. Fumio Hirata 60th birthday. Work of
Yu.H. was supported in part by the Austrian Fonds zur F\"orderung der
wi\-ssen\-schaft\-li\-chen Forschung under Project P 19583 and work of
J.I. by DFG grant NE410/8-2.


\label{last@page}
\end{document}